\documentclass[a4paper, 11pt]{article}

\usepackage{calc, amsmath, amsthm, a4, latexsym, amssymb, color, url, soul,dsfont, lineno, mathabx, dsfont}
\usepackage{graphicx}
\definecolor{comcolor}{rgb}{0.9,0.3,0.3}
\definecolor{starcolor}{rgb}{0.3,0.3,0.9}
\definecolor{hscolor}{rgb}{0.9,0.6,0.5}
\definecolor{darkgreen}{rgb}{0.1,0.6,0.3}

\newtheorem{thm}{Theorem}[section]

\newtheorem{prop}[thm]{Proposition}

\theoremstyle{definition}

\newtheorem{rem}[thm]{Remark}

\newcommand{\be}[1]{\begin{equation}\label{#1}}
\newcommand{\ee}{\end{equation}}
\newcommand{\ba}{\begin{array}}
\newcommand{\ea}{\end{array}}
\newcommand{\bal}{\begin{aligned}}
\newcommand{\eal}{\end{aligned}}

\newcommand{\N}{\mathbb{N}}
\newcommand{\Q}{\mathbb{Q}}
\newcommand{\Z}{\mathbb{Z}}
\newcommand{\E}{\mathbb{E}}

\renewcommand{\P}{\mathbb{P}}

\newcommand{\n}{\mathbf{n}}
\newcommand{\e}{\mathbf{e}}
\renewcommand{\t}{\mathbf{t}}
\newcommand{\tK}{\tt K}
\newcommand{\tW}{\tt W}
\newcommand{\tS}{\tt S}
\newcommand{\tTI}{\tt TI}
\newcommand{\tI}{{\tt I}}

\renewcommand{\t}{\mathbf{t}}

\renewcommand{\P}{\mathbb{P}}
\renewcommand{\H}{\mathcal{H}}

\begin{document}

\begin{center}
{\Large \bf Statistical tools for seed bank detection}\\[20mm]
\title{Statistical tools for seed bank detection}
{\large \bf Jochen Blath\footnote{Institut f\"ur Mathematik, Technische Universit\"at Berlin, Stra\ss e des 17.~Juni 136, 10623 Berlin, Germany; blath@math.tu-berlin.de}, Eugenio Buzzoni\footnote{Institut f\"ur Mathematik, Technische Universit\"at Berlin, Stra\ss e des 17.~Juni 136, 10623 Berlin, Germany; buzzoni@tu-berlin.de}, Jere Koskela\footnote{Corresponding author. Department of Statistics, University of Warwick, Coventry CV4 7AL, UK; j.koskela@warwick.ac.uk}, Maite Wilke-Berenguer\footnote{Fakult\"at f\"ur Mathematik, Ruhr-Universit\"at Bochum, Universit\"atsstra\ss e 150, 44801 Bochum, Germany; maite.wilkeberenguer@ruhr-uni-bochum.de}\\[20mm]}
\author{Jochen Blath\footnote{TU Berlin; blath@math.tu-berlin.de}, Eugenio Buzzoni\footnote{TU Berlin; buzzoni@tu-berlin.de}, Jere Koskela\footnote{Corresponding author. University of Warwick; j.koskela@warwick.ac.uk}, Maite Wilke-Berenguer\footnote{Ruhr-Universit\"at Bochum; maite.wilkeberenguer@ruhr-uni-bochum.de}}

\begin{abstract}
We derive statistical tools to analyze the patterns of genetic variability produced by models related to seed banks; in particular the Kingman coalescent, its time-changed counterpart describing so-called weak seed banks, the strong seed bank coalescent, and the two-island structured coalescent. 
As (strong) seed banks stratify a population, we expect them to produce a signal comparable to population structure.
We present tractable formulas for Wright's $F_{ST}$ and the expected site frequency spectrum for these models, and show that they can distinguish between some models for certain ranges of parameters.
We then use pseudo-marginal MCMC to show that the full likelihood can reliably distinguish between all models in the presence of parameter uncertainty under moderate stratification, and point out statistical pitfalls arising from stratification that is either too strong or too weak.
We further show that it is possible to infer parameters, and in particular determine whether mutation is taking place in the (strong) seed bank.
\end{abstract}
\end{center}


\newpage
\section{Introduction and basic models}
\label{sn:thebasicmodel}

\subsection{Seed banks in population genetics}

Seed banks, or reservoirs of dormant individuals that can be resuscitated in the future, are common in many communities of macroscopic (e.g.\ plant) and microscopic (e.g.\ bacterial) organisms. 
They extend the persistence of genotypes and are important for the diversity and functioning of populations.
Microbial dormancy is common in a range of ecosystems, and there is evidence that the ecology and evolution of microbial communities are strongly influenced by seed banks. 
It has been observed that more that 90\% of microbial biomass in soil is metabolically inactive. 
See \cite{LJ11, SL18} for overviews on seed banks.

Seed banks have a significant influence on classical evolutionary forces such as selection and genetic drift. 
For example, seed banks can counteract the effect of genetic drift, and lead to population stratification.
However, the development of a comprehensive population genetic theory incorporating seed banks is still in its early stages, and plenty of open questions remain \cite{SL18}.
While some basic mathematical models have been derived and predict unique patterns of genetic variability in idealized scenarios \cite{LJ11, KKL01, ZT12, BEGCKW15, BGCKW16, dHP17, KMTZ17, HMTZ18}, statistical tools to infer the presence of `weak' or `strong' seed banks are still largely missing (however, see \cite{SAMT19}, which was produced in parallel with this work).

The aim of this article is to provide basic statistical tools to analyze patterns of genetic variability produced by the above models of seed banks.
We also assess the utility of these tools for parameter estimation and model selection based on genetic data. 
Notably, we will provide comparisons between variability under seed banks, and classical models of population structure \cite{H94}.
Both model classes can be expected to predict somewhat similar patterns of diversity, and we will study the extent to which sequence data can differentiate between. This is motivated by the need to understand the roles of dormancy and biogeography in microbial communities \cite[p.125]{LJ11}.
We extend earlier studies \cite{T11, BEGCKW15}, where seed banks were compared to panmictic models.
We begin with a brief review of the relevant genetic models with and without seed banks.

\subsection{Population models}

\emph{Kingman's coalescent} ($\tK$):
The standard model of genetic ancestry in the absence of a seed bank is the \emph{coalescent} (or {\em Kingman's coalescent}) \cite{K82}, which describes ancestries of samples of size $n \in \N$ from a large, selectively neutral, panmictic population of size $N \gg n$ following e.g.\ a Wright-Fisher model. 
Measuring time in units of $N$ and tracing the ancestry of a sample of size $n \ll N$ backwards in time results in a coalescent process $\Pi^n$ in which each pair of lineages merges to a common ancestor independently at rate 1 as $N \rightarrow \infty$. 
A rooted ancestral tree is formed once the most recent common ancestor of the whole sample is reached.
We denote this scenario by $\tK$. 
This model is currently the standard null model in population genetics (see e.g.\ \cite{W08} for an introduction) and arises from a large class of population models. 

\emph{`Weak' seed banks and the delayed coalescent} ($\tW$):
The coalescent was extended in \cite{KKL01} to incorporate a \emph{`weak' seed bank}. 
In this model, an individual inherits its genetic material from a parent that was alive a random number of generations ago. 
The random separation is assumed to have mean $\beta^{-1}$ for some $\beta \in ( 0, 1 ]$. 
Measuring time in units of $N$ and tracing the ancestry of a sample of size $n \ll N$ as above, it can be shown that the genealogy is still given by a coalescent in which each pair of lineages merges to a common ancestor independently with rate $\beta^2$.
Thus, the effect of the seed bank is to stretch the branches of the Kingman coalescent by a constant factor \cite{KKL01, BGCKS12}, but the topology and relative branch lengths remain identical to those of the coalescent.
Thus the weak seed bank coalescent with mean separation $\beta^{-1}$ and population-rescaled mutation rate $u > 0$ is statistically identical to Kingman's coalescent with population-rescaled mutation rate $u / \beta^2$, and e.g.~the normalized site frequency spectrum under the infinitely many sites model is invariant between these models \cite{BEGCKW15}.
Nevertheless, the seed bank does have important consequences e.g.\ for the estimation of effective population size and mutation rates in the presence of prior information, or some other means of resolving the lack of identifiability.
We call the corresponding coalescent a `delayed coalescent' and denote this scenario by $\tW$. 

\emph{`Strong' seed banks and the seed bank coalescent} ($\tS$): 
The recent model in \cite{BGCKW16} extends the Wright Fisher framework to a model with a classical `active' population of size $N$ and a separate `seed bank' of comparable size $M:=\lfloor N / K \rfloor$, for some $K > 0$, allowing for `migration' of a fraction of $\lfloor c / N\rfloor$ individuals between the two subpopulations. 
The active population follows a Wright-Fisher model, while the dormant population in the seed bank persists without reproducing.
This model can be seen as a mathematical formalization of \cite[Figure 2]{LJ11}. 
The age structure in the resulting seed bank is geometric with mean of order $N$, which means that seeds can remain viable in the seed bank for $O(N)$ generations. 
Measuring time in units of $N$, the genealogy of a sample of size $n^{(1)} \ll N$ (resp.\ $n^{(2)} \ll M$) from the active (resp.\ dormant) population, is described by the so-called {\em seed bank coalescent} \cite{BGCKW16}, in which active lineages fall dormant at rate $c$ and coalesce at rate 1 per pair, while dormant lines resuscitate at rate $cK$.
We call this ancestral process a \emph{(strong) seed bank coalescent}, and denote this scenario by $\tS$.
The seed bank coalescent has a very different site frequency spectrum to the classical and weak seed bank coalescents \cite{BEGCKW15}. 

\emph{The two island model and the structured coalescent ($\tTI$)}:
Having modeled a strong seed bank as a separate population linked to the active one via migration, it is natural investigate its relation to Wright's two island model \cite{H94, W08}. 
In the simplest case (which we assume throughout) there are two populations (1 and 2) of respective sizes $N$ and $M = \lfloor N / K \rfloor$, with a fixed fraction of $\lfloor c / N \rfloor$ individuals migrating both from 1 to 2 and from 2 to 1 each generation.
Measuring time in units of $N \rightarrow \infty$ generations, the genealogy of a sample of respective sizes $n^{(1)} \ll N$ and $n^{(2)} \ll M$ from islands 1 and 2 is described by a similar ancestral process as the strong seed bank coalescent, except that pairs of lineages in population 2 also merge independently with rate $1 / K$.
We denote this scenario by $\tTI$. 
The resulting ancestral process is the \emph{structured coalescent} \cite{H94, Not90}, which describes the ancestry of a geographically structured population with migration.

In this article we investigate the extent to which genetic data can distinguish between models $\tK$, $\tW$, $\tS$, and $\tTI$. 
All four are a priori plausible as models for various real populations.
In \cite{T11}, the authors studied two species of wild tomato (S.~chilense and S.~peruvianum), and inferred average seed bank delays of 9 and 12 generations.
Estimates of corresponding effective population sizes are $O(10^5)$ \cite{A7}, which suggests that scenario $\tW$ is appropriate.
On the other hand, dormant bacteria have been observed to remain viable for millions of years \cite{V00}, which suggests that the strong seed bank could be relevant. 
A stable reservoir of dormant individuals requires periods of dormancy on the order of the effective population size \cite{BEGCKW15}, so that model $\tS$ seems appropriate whenever there is a stable reservoir of dormant types, with individuals switching between reservoirs with some fixed rate as outlined in \cite{LJ11} for bacterial communities.
These considerations highlight the need to distinguish the two types of seed banks from data in cases where the presence or size of a seed bank or the typical period of dormancy are uncertain.
It is also of interest to distinguish the signal of (strong) seed banks from geographic structure, which could in principle produce similar patterns of genetic stratification in the population. 

\subsection{Mutation models and key statistical quantities} \label{subsec:mut_models}

We consider three models of genetic diversity and mutation: the infinite alleles model (IAM), the infinite sites model (ISM), and the finite alleles model (FAM).
The FAM is in less widespread use due to its high computational demands, so we postpone results under it to the appendix.
We also only present results under the FAM for the special case of two alleles, but our work generalizes to any number.

We consider several classical statistical quantities: the sample heterozygosity and Wright's $F_{ST}$ \cite{W51}, the site frequency spectrum (SFS), and the sampling distribution of the full sequence data.
These measures are informative about the underlying coalescent scenario, and suited to the different mutation models, to varying degrees.
They also differ in the extent to which they are tractable. 
The sample heterozygosity, Wright's $F_{ST}$ and the (normalized) SFS discard statistical signal, but are readily computed (at least numerically) in most settings. 
The sampling distribution of the sequence data fully captures the signal in a data set, but is available only via Monte Carlo schemes.
Our results clarify when computationally cheap summary statistics suffice to distinguish between models, and when the full likelihood is needed.

\emph{The infinite alleles model (IAM)}: Given a coalescent tree distributed according to any of the models introduced above, a sample of genetic data from the infinite alleles model is generated by assigning an arbitrary allele to the most recent common ancestor, and simulating mutations along the branches of the coalescent tree with population-rescaled mutation rate $u := N \mu > 0$ for the branches in the first (and possibly only) population and $u' := M \mu' \geq0$ in the second population (if one is present).
Above, $N$ and $M$ represent effective population sizes, while $\mu$ and $\mu'$ are the per-site, per-generation mutation probabilities.
Each mutation results in a new parent-independent allele that has never existed in the population before, and alleles are inherited along lineages.
Population-rescaled mutation rates in further mutation models below are defined analogously.

We encode a sample of size $n^{(1)} + n^{(2)} = n$, where $n^{(i)}$ is the sample size from population $i$, as the pair of $n$-tuples $( \n^{(1)}, \n^{(2)} )$, where $n_j^{(i)}$ is the number of $j$ alleles on island $i$ under some fixed but arbitrary ordering of observed alleles, and $n^{ ( i ) } = \sum_j n_j^{ ( i ) }$.
Both tuples are padded by zeros if fewer than $n$ distinct alleles are observed for notational convenience, and we will drop the superscripts and second tuple for models with only one population.

The (somewhat out-dated) infinite alleles model is appropriate when the data only encodes when two alleles are different, but no further detail is available, such as is the case for electrophoresis data \cite{H66}. 

\emph{The infinite sites model (ISM)}:
We now identify the locus with the unit interval $[0, 1]$.
Mutations, which continue to occur on the branches of the coalescent tree with rates $u$ and $u'$, are assumed to occur at distinct sites on the locus, and are inherited along the branches of the tree so that the allele of an individual is the list of all mutations along its ancestral line. 
Thus, the whole history of mutations up to the root is retained. 
A sample of size $n := n^{(1)} + n^{(2)}$ is specified by the triple $(\t, \n^{(1)}, \n^{(2)})$, where $\t := (t_1, \ldots, t_d)$ is the list of all observed alleles, and $n_j^{(i)}$ is the observed number of allele $t_j$ in population $i$.  
For details on this parametrization of the infinite sites model and its relation to coalescent models see e.g.\ \cite{BB08}.

\emph{The finite alleles model (FAM)}: 
We consider a finite set of possible allele identified with $\{1, \ldots, d\}$. 
The type of the most recent common ancestor is sampled from some probability mass function $\rho = (\rho_1, \ldots, \rho_d)$, and mutations occur along the branches of the coalescent tree at rates $u$ and $u'$ as before.
At a mutation, a new allele is sampled from a $d \times d$ stochastic matrix $P$, and alleles are inherited along branches as before.
A sample under the FAM is also described by the pair of tuples $( \n^{(1)}, \n^{(2)} )$, with the distinction that each tuple is now of fixed length $d$.
Throughout the article, we take $d = 2$, and set $u_2 := u P_{1 2}$ as well as $u_1 := u P_{2 1}$ for notational brevity, and define mutation rates $u'_1$ and $u'_2$ for a second population analogously.

Note that the classical Watterson estimator of mutation rate depends on the chosen coalescent model.
Further, in scenarios $\tTI$ and $\tS$, we will allow the overall mutation rate to differ between active and dormant lineages.
Determining whether mutations take place on dormant lineages in nature, perhaps at a reduced rate, is an interesting open question \cite{SL18}, and one of our motivations was to determine whether it is answerable from DNA sequence data. 

\subsection{Diffusion models}
All four coalescent models are dual to their respective {\em Wright-Fisher diffusions}, the exact form of which depends on the accompanying mutation model. 
The FAM, $\tTI$ Wright-Fisher diffusion solves the pair of SDEs
\begin{align}
\label{eq:system_general}
\text{d} X(t) & = [ u_2 (1 - X(t)) - u_1 X(t) + c (Y(t) - X(t)) ] \text{d}t \notag \\
&\phantom{=}+ \alpha \sqrt{X(t)(1-X(t))}\text{d}B(t), \notag \\
\text{d} Y(t) & = [ u_2' (1 - Y(t)) - u_1' Y(t)  + K c (X(t) -Y(t)) ] \text{d}t \notag \\
&\phantom{=} + \alpha' \sqrt{Y(t) (1-Y(t))} \text{d}B'(t), 
\end{align}
with initial value $(X(0), Y(0)) =(x,y) \in [0,1]^2$, where $1 / \alpha^2$, $1 / ( \alpha' )^2$ are effective population sizes, and $\{B_t\}$, $\{B'_t\}$ are independent Brownian motions.
Duals to scenarios $\tK$, $\tW$, and $\tS$ can be recovered as special cases: for $\tK$ we set $\alpha = 1$ and $c =0$, for $\tW$ we take $\alpha = \beta$ and $c = 0$, and for $\tS$ we take $\alpha = 1$ and $\alpha' = 0$.
For scenarios $\tK$ and $\tW$ we also only consider the $X$-coordinate, and in scenario $\tS$, the $X$-coordinate corresponds to the active population, while $Y$ is the seed bank.
In each case the solution is an ergodic diffusion with a unique stationary distribution on $[0, 1]$ (or $[0, 1]^2$), which we will denote by $\mu^{\tI}$ for $I \in \{ \tK, \tW, \tS, \tTI \}$.
It is also possible to derive the analogue of the Wright-Fisher diffusion for the IAM and ISM. 
This leads to measure-valued diffusions, or {\em Fleming-Viot processes} \cite{EK86}, which we do not require in our analysis.

\subsection{Outline of the paper} 

In Section \ref{classical_measures} we discuss Wright's $F_{ST}$ and the site frequency spectrum (SFS). 
We use phase-type distribution methods \cite{HSJB18} to compute the expected SFS, and show that these statistics can distinguish between our scenarios to some extent.
Since they are cheap to compute, they serve as a plausibility check for the presence of seed banks.
Results for Wright's $F_{ST}$ for the FAM are presented in the appendix.

In Section \ref{sampling_distributions} we present recursions for the likelihood functions of observed sequence data for the IAM and ISM under scenario $\tS$, which are currently missing in the literature. 
The recursions are intractable for large sample sizes, so we provide low-variance importance sampling schemes to approximate their solutions.
Corresponding results for the FAM are presented in the appendix.

In Section \ref{model_selection} we provide statistical machinery for model selection and parameter inference for all scenarios under the ISM, which is the most relevant for handling of real data.  
We employ a pseudo-marginal Metropolis-Hastings algorithm for simultaneous model selection and parameter inference for the different models and assess its effectiveness with simulated data sets. 
We also address the specific question of detecting mutation in the (strong) seed bank.

We conclude the paper with a discussion of our results in Section \ref{sec:discussion}.

\section{Classical measures of population structure}
\label{classical_measures}

In this section we investigate classical summary statistics for inferring population structure, namely Wright's $F_{ST}$ and the (normalized) site frequency spectrum nSFS.
Unless stated otherwise, we assume positive mutation rates in all (sub-)populations.

\subsection{Wright's \texorpdfstring{$F_{ST}$}{FST} for seed banks and structured populations}
\label{ssn:FST}

Wright's $F_{ST}$ \cite{W51} is a prominent but crude measure for population structure.
There are various (more-or-less equivalent) formulations in the literature.
Here, we follow the notation and interpretation of Herbots \cite[p.\ 73]{H94} (see also \cite[Chapter 3]{R04}), which studies this quantity for various structured  models. 
Define
\begin{align}
\label{eq:FST_def}
F_{ST}:= \frac{p_0-\bar p}{1-\bar p},
\end{align}
where $\bar p$ is the probability of \emph{identity} of two genes sampled uniformly at random from the whole population, while $p_0$ is the probability of identity of two genes sampled uniformly from a single sub-population, itself previously randomly sampled with probability given by its relative population size. 

For the FAM, $\bar p$ and $p_0$ are determined by the \emph{sample homozygosity} (discussed in the appendix), whereas for the IAM and ISM, they are given in terms of {\em identity by descent}.
Positive values of $F_{ST}$ indicate population structure, though its exact interpretation depends on the biological scenario. Hartl and Clark argue that $F_{ST} \in (0.05, 0.15)$ constitutes ``moderate'' genetic differentiation \cite[Section 6.2]{H97}, though applying such a rule indiscriminately can be misleading as $F_{ST}$ values depend on e.g.~the life cycle and reproductive characteristics of the species, as well as on the details of spatial structure.
We will be interested how the quantity compares between $\tS$ and $\tTI$, where the latter certainly represents a strongly structured population.

\paragraph{Wright's $F_{ST}$ for the IAM}

Under the IAM, every mutation leads to a distinct allele.
Hence, two sampled individuals are identical if and only if neither of their ancestral lineages mutated since the time of their most recent ancestor. 
Thus $p_0$ and $\bar p$ from \eqref{eq:FST_def} can be expressed as the so-called probabilities of {\em identity by descent} (IBD), and these probabilities can easily be represented in terms of the relevant coalescent.

Let $T$ be the (random) \emph{time to the most recent common ancestor} (TMRCA) of a sample of size 2 in any of the above coalescent models and observe that, if we assume the same mutation rate $u=u'$ in both sub-populations (for $\tS, \tTI$), the probability that we do not see any mutations along the branches of the coalescent up to a time $t>0$ is given by $e^{-2 u t}$. 
Since mutations occur conditionally independently given $T$, we have 
\begin{align*}
p_0 = \E_{\pi_0}[ e^{ -2 u T } ] \qquad \text{and} \qquad \bar p =\E_{\bar \pi}[ e^{-2 u T} ],
\end{align*}
where 
 \begin{align*}
  \pi_0 := \left(\frac{K}{1+K},\,0\,,\,\frac{1}{1+K},\,\right), \qquad  \bar\pi := \left(\frac{K^2}{(1+K)^2},\frac{2K}{(1+K)^2},\,\frac{1}{(1+K)^2}\,\right).
 \end{align*}
In words, $\E_{\pi_0}$ is the expectation when the both genes are sampled from the same population, itself previously sampled among all populations according to its relative size, and $\E_{\bar \pi}$ is the expectation when the genes are sampled uniformly from the whole population. 
IBD has recently been investigated for $\tS$ in \cite{dHP17} in the case of a finite population with seed bank on a discrete torus. 

To obtain an expression for IBD for distinct mutation rates $u\neq u'$, we need to trace the time the lineages spend in each population before the TMRCA. 
Let $R_{2,0}$, $R_{1,1}$ and $R_{0,2}$ be the time until coalescence the ancestral lineages spend both in the first population, one lineage in each population and both in the second population, respectively. 
Then $T=R_{2,0}+R_{1,1}+R_{0,2}$ and we get
\begin{align*}
p_0 &= \E_{\pi_0}\left[ e^{ -2 u R_{2,0} - (u+u')R_{1,1} - 2u'R_{0,2} } \right], \\
\bar p &=\E_{\bar \pi}\left[ e^{ -2 u R_{2,0} - (u+u')R_{1,1} - 2u'R_{0,2} } \right].
\end{align*}

\begin{figure}[!h]
\centering
\includegraphics[width = 0.49\textwidth]{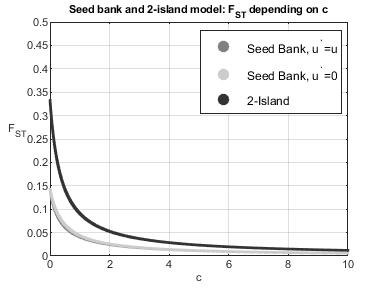}
\includegraphics[width = 0.49\textwidth]{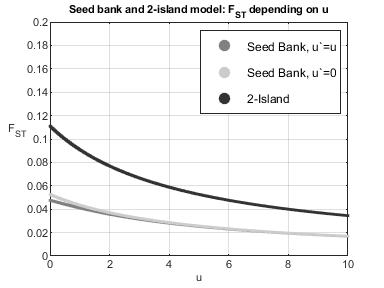}
\includegraphics[width = 0.49\textwidth]{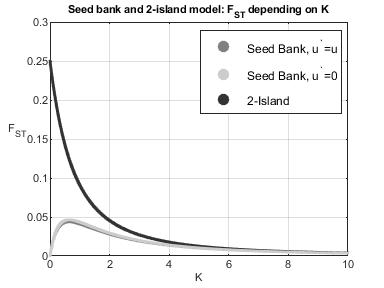}
\caption{$F_{ST}$ under $\tS$ and $\tTI$ as a function of various parameters in the IAM. Where not specified, $K=c=1$, $u_1=u_2=0.5$.}
\label{fig:7}
\end{figure}

Phase-type distribution theory \cite{HSJB18} yields elegant closed form expressions for these quantities.
\begin{prop}\label{prop:Fst}
 Assuming the IAM, the fixation index $F^{\tI}_{ST}$ for $\tI \in \{\tS,\tTI\}$ is given by 
 \begin{align*}
  F^{\tI}_{ST} = \frac{p^{\tI}_0 - \bar p^{\tI}}{1-\bar p^{\tI}}
 \end{align*}
 where
 \begin{align*}
  p^{\tI}_0 	 = \pi_0(A-S^{\tI})^{-1}s^{\tI} \qquad \text{and} \qquad   \bar p^{\tI} = \bar \pi (A-S^{\tI})^{-1}s^{\tI} 
 \end{align*}
where $A$ is a diagonal matrix with diagonal $[ -2 u, -( u + u' ), -2 u' ]$, and 
 \begin{align*}
  S^{\tI} = \begin{bmatrix}
            -(2c + 1) & 2c & 0 \\
             cK & -(cK+c) & c \\
             0 & 2cK & -(2cK + \alpha^{\tI}) 
          \end{bmatrix} 
          \quad \text{and} \quad  s^{\tI}=  \begin{bmatrix}
                                  1\\ 0 \\ \alpha^{\tI}
                                 \end{bmatrix},
 \end{align*}
where $\alpha^{\tS} = 0$ and $\alpha^{ \tTI} = 1 / K$.
\end{prop}
The proof is obtained using the machinery of \cite{HSJB18} and we adhere to the notation used therein for the convenience of the reader. 
See \cite[Example 2.4]{HSJB18} for some different functionals of the seed bank coalescent obtained in this way.
\begin{proof}
 Let $\{Z_t\}$ be a time-continuous Markov chain on the finite space 
 \begin{align*}
  E_2 :=\{(2,0),(1,1),(0,2), (*,*)\}  
 \end{align*}
 with Q-matrix
 \begin{align*}
  Q^{\tI} = \begin{bmatrix}
       S^{\tI} & s^{\tI}\\
       0 & 0
      \end{bmatrix} 
 \end{align*}
for $\tI \in \{ \tS, \tTI \}$.
For each model, $\{Z_t\}$ traces whether the lineages of a sample of 2 are both in the first population, one in each population or both in the second population. 
The state $(*,*)$ is reached at time $T$, and is absorbing.

Recall that $R_{2,0}$ was the time the ancestral lineages of the sample spent both in the first population and note that we can write it as 
 \begin{align*}
  R_{2,0} = \int_0^T \mathds{1}_{ \{ ( 2, 0 ) \} }(Z_t)\text{d}t.
 \end{align*}
We can do the same for $R_{1,1}$ and $R_{0,2}$, and thus \cite[Theorem 2.5]{HSJB18} yields
 \begin{align*}
  p_0	& =  \E_{\pi_0}\left[ e^{ -2 u R_{2,0} - (u+u')R_{1,1} - 2u'R_{0,2} } \right]\\
	& = \pi_0 \left(\begin{bmatrix}
	                 -2u & 0 & 0 \\
			  0 & -(u+u') & 0 \\
			  0 & 0 & -2u'
	                \end{bmatrix} - S^{\tI}\right)^{-1}s^{\tI}
 \end{align*}
and analogously for $\bar p$. 
\end{proof}
Figure \ref{fig:7} illustrates the $F_{ST}$ under different choices of parameters for the IAM. 
The pictures differ only slightly from those of the FAM in Figure \ref{fig:1} in the appendix. 

\paragraph{Wright's $F_{ST}$ for the ISM}
The central difference between the IAM and the ISM is that all previous mutations on a lineage remain observable in the latter.
However, this does not affect the probability of IBD of two individuals --- they will still carry the same allele if and only if neither ancestral line mutated between the TMRCA and the present.
Thus, sample heterozygosity $H$ and $F_{ST}$ under the ISM can be computed in exactly the same way as in the IAM.

\subsection{The site frequency spectrum (SFS) in the ISM}

The SFS is one of the most frequently used summary statistics under the ISM.
For a sample of size $k$ it is given by a vector $(\zeta_1^{(k)},\dots, \zeta_{k - 1}^{(k)})$, with $\zeta_i^{(k)}$ denoting the number of sites at which the \emph{derived} allele is observed $i$ times in the sample. 
This assumes that we know the wildtype and are therefore able to determine which of the two alleles is derived, and which is ancestral.
In the case where we do not know which allele is which, the \emph{folded} SFS $(\eta_1^{(k)},\dots, \eta_{ \lfloor k / 2 \rfloor }^{(k)})$ can be used instead, where $\eta_i^{(k)}$ is the number of sites where two alleles are observed with multiplicities $i : k - i$.

The SFS is well understood for the classical Kingman coalescent ${\tK}$, and thus also in the case ${\tW}$, since the weak seed bank coalescent is just a constant time-change of the Kingman coalescent \cite[Formula 1]{ZT12}.

We can also calculate the \emph{expected} SFS for the cases $\tTI$ and $\tS$. We consider $k$ individuals sampled according to some initial distribution $\pi$ from the first and the second population. 
Since mutations in the ISM occur according to a Poisson process conditionally on the coalescent, $\E_{\pi}[\zeta^{(k)}_i]$ is the product of the mutation rate and the expected total lengths of branches that are ancestral to $i$ individuals, for which phase-type distribution theory is well suited. 
In order to state the result (and thereby give the bulk of the proof), we require a few technical definitions, but the calculation of the SFS then reduces to a simple vector-matrix multiplication in Proposition \ref{prop:SFS}. 
The structure is reminiscent of the observations for the SFS of $\Lambda$-coalescents in \cite{HSJB18}. 

As in Proposition \ref{prop:Fst} we want to define an auxiliary Markov chain. Its state space $E$ should be small to minimize computational cost, but needs to be sufficiently large to contain all information necessary to calculate the SFS, i.e.\ we need to know how many lineages are ancestral to $i$ individuals in the sample at any time in the coalescent, and how many of these lineages are in the first and second populations, respectively, in order to account for different mutation rates. 
For a sample of size $k$ define 
\begin{align*}
 E:=\left\{ a \in \{0, \ldots, k\}^{2k} \; \Big| \; \sum_{i=1}^ki(a_i+a_{k+i})=k\right\}\setminus \{e_k, e_{2k}\}
\end{align*}
where $e_k$ and $e_{2k}$ are the vectors with the entry 1 in positions $k$ and $2k$ respectively (and thus $0$ everywhere else). We remove these in order to identify them as what will be the unique absorbing state of the Markov chain. Thus define
\begin{align*}
 E^*:=E\cup \{\bigast\}.
\end{align*}

For $a \in E$, if $i = 1, \ldots k$, the quantity $a_i$ is the number of lineages currently in the first population that are ancestral to $i$ of the sampled individuals (independently of their origin). If $i=k+1, \ldots, 2k$ then $a_i$ is the analogous number of lineages currently in the second population. 

Given this interpretation, it becomes easy to identify the set $E_0$ of sensible starting points for the auxiliary Markov chain:
\begin{align*}
 E_0:=	& \{a \in E \mid a_1+a_{k + 1}=k\}.
\end{align*}
Starting in $a \in E_0$ corresponds to a sample of $a_1$ individuals from the first and $a_{k+1}$ individuals from the second population.
 Let $\pi$ be the initial a distribution of the Markov chain, assumed concentrated on $E_0$. 

The only allowed transitions of the chain will be those corresponding to a coalescence or a migration.
For $z \in \Z$ let $(z)^+:=\max\{z,0\}$ and $(z)^-:=\min\{z,0\}$. 
We call a transition from the state $a \in E$ to $b \in E$ a \emph{coalescence} if
\begin{enumerate}
 \item $\sum_{j=1}^{2k} (b_j-a_j)^- = -2$,
 \item $\sum_{j=1}^{2k} (b_j-a_j)^+ = 1$,
 \item $\sum_{j=1}^k j(b_j-a_j) = 0$.
\end{enumerate}
The first two describe the effect of the coalescence of two lineages. 
The last sum only runs until $k$, ensuring that the coalescence takes place between lineages in the same population.
A transition from $a$ to $b$ will be called a \emph{migration} if 
\begin{enumerate}
 \item $\sum_{j=1}^{2k} (b_j-a_j)^- = -1$,
 \item $\sum_{j=1}^{2k} (b_j-a_j)^+ = 1$.
\end{enumerate}
The rates at which the Markov chain then transitions between the states $a, b\in E$ depend on the model and are given by 
\begin{align*}
 S^{\tI, c}_{a,b}	& :=  \prod_{j=1:\, b_j-a_j<0}^k\binom{a_j}{b_j-a_j} + \alpha^{ \tI } \prod_{j=1:\, b_{k+j}-a_{k+j}<0}^k\binom{a_{k+j}}{b_{k+j}-a_{k+j}}, 
\end{align*}
if $a\mapsto b$ is a  \emph{coalescence}  and 
\begin{align*}
 S^{\tI, m}_{a,b}	& := c\sum_{j=1:\, b_j-a_j<0}^k a_j + cK \sum_{j=1:\, b_{k+j}-a_{k+j} <0}^ka_{k+j},
\end{align*}
if it is a \emph{migration}, where we again set $\alpha^{ \tI } = 0$ if $\tI = \tS$ and $\alpha^{ \tI } = 1/K$ if $\tI=\tTI$. 

Next, define $s^{\tI}: E \rightarrow [0, \infty[$ as
\begin{align*}
 s^{\tI}(a):= \begin{cases}
               1, 		& \text{if } \sum_{j=1}^{2k}a_j=\sum_{j=1}^{k}a_j = 2,\\
               \alpha^{ \tI },		& \text{if } \sum_{j=1}^{2k}a_j=\sum_{j=k+1}^{2k}a_j = 2,\\
               0,		& \text{otherwise.}
              \end{cases}
\end{align*}
Note that $s^{\tI}$ is non-zero precisely on the states with two lineages remaining which could coalesce into the absorbing state $\bigast$, and gives the rate of that event.

With this now define the matrix $S^{\tI} = (S^{\tI}_{a,b})_{a,b \in E}$ through
\begin{align*}
 S^{\tI}_{a,b} : = \begin{cases}
		      S_{a,b}^{ \tI, c },				&\text{if } a \mapsto b \text{ is a coalescence},\\		      		      S_{a,b}^{ \tI, m },				&\text{if } a \mapsto b \text{ is a migration},\\
		      - s^{\tI}(a) - \sum_{a' \neq a } S_{a,a'}	&\text{if }a=b, \\
		      0,				&\text{otherwise }.
                   \end{cases}
\end{align*}

Finally, we define $r_i(\bigast):=0$ for any $i = 1, \ldots, k-1$, and for every $a \in E$,
\begin{align*}
 r_i(a)		& := ua_i + u'a_{k+i}. 
\end{align*}

The vectors $\pi$, $r_1, \ldots, r_{k-1}$ can be taken as normal vectors by fixing an ordering on $E^*$, which also justifies representing $S^{\tI}$ as a matrix.
Hence \eqref{eq:SFS} should be read as a vector-matrix multiplication.

\begin{prop}\label{prop:SFS}
 Assume the ISM, with mutation rates $u, u'\geq 0$ in the first and second population, respectively. 
 Let $\pi$ describe how the $k\in \N$ individuals are sampled from the first and second population. 
Then
 \begin{align}\label{eq:SFS}
  \E_{\pi}\left[\zeta^{(k)}_i\right] = \pi(-S^{\tI})^{-1}r_i
 \end{align}
for all $i = 1, \dots, k-1$ and $\tI \in \{\tTI, \tS\}$.
\end{prop}
For a sample of $k_1$ individuals from the first population and $k_2=k - k_1$ individuals from the second population, set $\pi = \pi^{ ( k_1, k_2 ) } := \delta_{ ( k_1, 0 \ldots, 0, k_2, 0, \ldots, 0 ) }$, where the right hand side is the Dirac delta measure and the non-zero entries are in positions 1 and $k + 1$.
For a sample drawn uniformly from the whole population, set $\pi( a ) = \pi^{ \text{unif} }( a ) := \binom{ k }{ a_{ k + 1 } } K^{ a_{ k + 1 } }{ ( K + 1 )^{-k} }$ for any $a \in E_0$.
\begin{proof}
 Let $\{Z_t\}$ be a Markov process with state space $E^*$ and Q-matrix 
 \begin{align*}
  Q:=\begin{bmatrix}
   S^{\tI}& s^{\tI} \\
   0		& 0
  \end{bmatrix}.
 \end{align*}
Started in $\pi$, the time $\{Z_t\}$ absorbs into $\bigast$ is equal in distribution to the time to the most recent common ancestor of a sample of size $k$ drawn according to $\pi$.
 Since mutations occur independently of the coalescent given the ancestry, to compute $\E_{\pi}[\zeta^{(k)}_i]$ we trace the time a lineage in the coalescent is ancestral to $i$ of the initial individuals and multiply it by $u$ when it is in the first and by $u'$ when it is in the second population.
 This is done by defining
 \begin{align*}
  \tilde \zeta^{(k)}_i := \int_0^{\tau} r_i(Z_t)\text{d}t,
 \end{align*}
and noting that
 \begin{align*}
  \E_{\pi}\left[\zeta^{(k)}_i\right] = \E_{\pi}\left[\tilde \zeta^{(k)}_i\right].
 \end{align*}
Thus,  \cite[Eq (8)]{HSJB18} yields \eqref{eq:SFS} above.
\end{proof}

\begin{rem}
The normalized expected site frequency spectrum \cite[p.\ 13]{EBBF15} (NESFS) $(E \hat{\zeta}_1^{(k)}, \dots, E \hat{\zeta}_{ k - 1 }^{(k)})$ is defined as
\begin{equation*}
 E \hat{\zeta}_i^{(k)} := \frac{\mathbb{E}[\zeta_i^{(k)}]}{\sum_{l=2}^{k} l \mathbb{E}[T_l]},
 \end{equation*}
where $T_l$ is the time during which there are $l$ distinct lineages in the coalescent regardless of to which population they belong. 
In other words, $\sum_{l=2}^k l \mathbb{E}[T_l]$ is the average tree length.
The NESFS is a first-order approximation of the expectation of the  normalized SFS \cite[p.\ 9]{EBBF15}, given by 
\begin{equation*}
 \hat{\zeta}_i^{(k)} := \frac{\zeta_i^{(k)}}{\zeta_1^{(k)} + \dots + \zeta_{k-1}^{(k)}}.
 \end{equation*}
 The distribution of $(\hat{\zeta}_{1}^{(k)}, \ldots, \hat{\zeta}_{k-1}^{(k)})$ is very insensitive to the mutation rate, provided it is not too small, facilitating practical inference when the mutation rate is unknown \cite[Supporting Information, pages SI12 -- SI13]{EBBF15}. 
 The average tree length for {\tS} was analyzed in \cite{HSJB18} and thus all necessary quantities to calculate the normalized expected SFS similarly to the SFS are given.
\end{rem}

Figures \ref{fig:10} and \ref{fig:12} provide illustrations of the expected SFS, with and without normalization. 
It is noteworthy that the magnitude of entries in the expected SFS varies strongly between the three models, while $\tS$ and $\tTI$ have very similar normalized spectra when only sampling one (in the case of $\tS$, active) population.
The implication is that all three models are straightforward to tell apart if the population-rescaled mutation rate is known, but that a larger sample or a more informative statistic is needed to distinguish $\tS$ from $\tTI$ when it is unknown.
When dormant lineages are included in the sample, $\tS$ predicts an excess of singletons, which can be detected even from the nSFS.
The excess of singletons is due to the fact that only few distinct active lineages will remain in the genealogy by the time that the ancestral line of the dormant samples first fell dormant.
Thus, the external branch connecting a dormant lineage to the rest of the ancetral tree is likely to be much longer than the external branches between active samples.

\begin{figure}
\centering
\includegraphics[width = 0.49\textwidth]{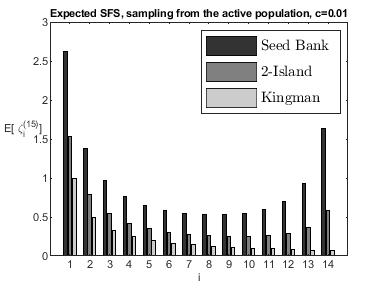}
\includegraphics[width = 0.49\textwidth]{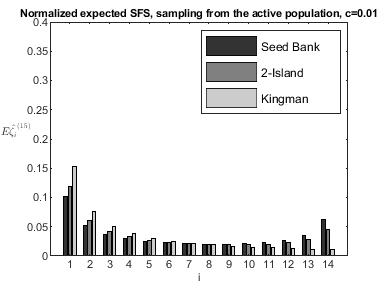}
\includegraphics[width = 0.49\textwidth]{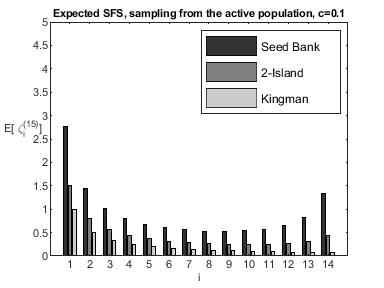}
\includegraphics[width = 0.49\textwidth]{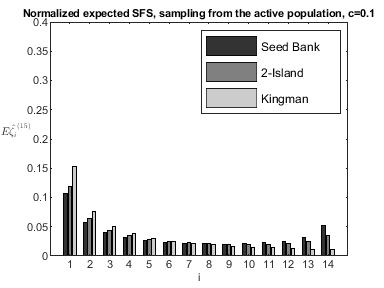}
\includegraphics[width = 0.49\textwidth]{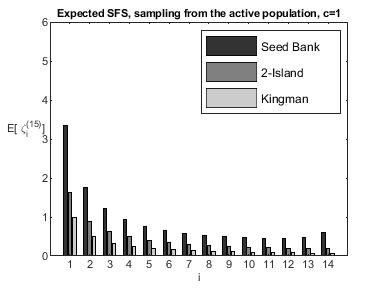}
\includegraphics[width = 0.49\textwidth]{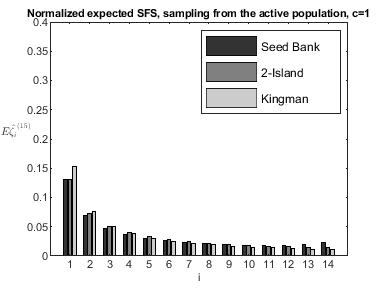}
\includegraphics[width = 0.49\textwidth]{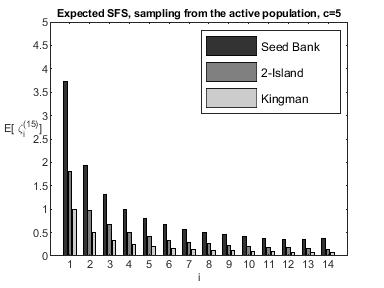}
\includegraphics[width = 0.49\textwidth]{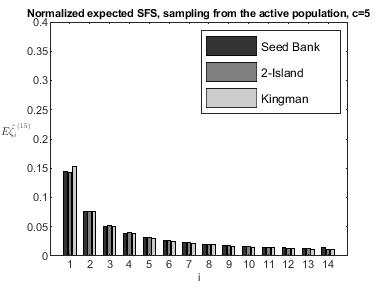}
\end{figure}
\begin{figure}
\centering
\includegraphics[width = 0.49\textwidth]{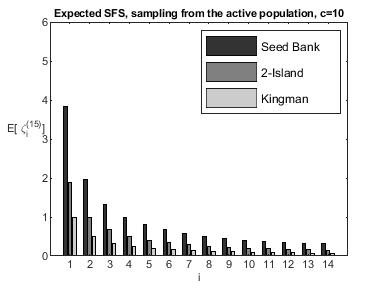}
\includegraphics[width = 0.49\textwidth]{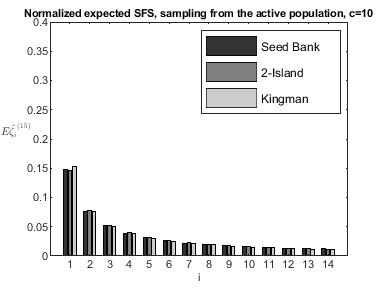}
\caption{Expected SFS sampled from the active population, i.e.\ $\pi^{(15,0)}$, with $K = u =1$.}
\label{fig:10}
\end{figure}

\begin{figure}[!h]
\centering
\includegraphics[width = 0.49\textwidth]{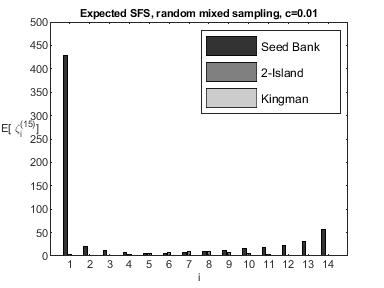}
\includegraphics[width = 0.49\textwidth]{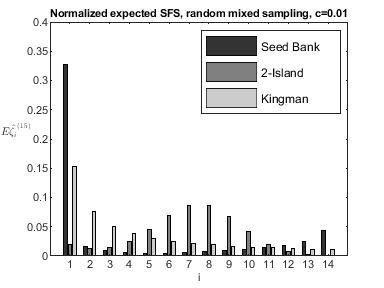}
\includegraphics[width = 0.49\textwidth]{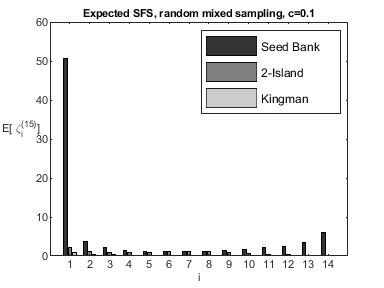}
\includegraphics[width = 0.49\textwidth]{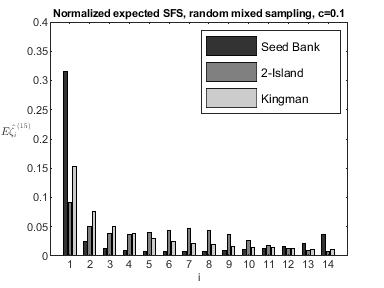}
\includegraphics[width = 0.49\textwidth]{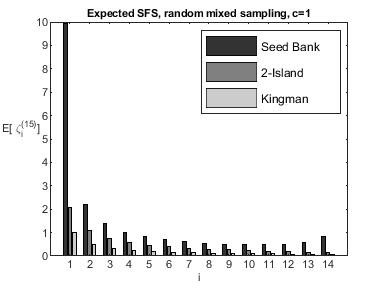}
\includegraphics[width = 0.49\textwidth]{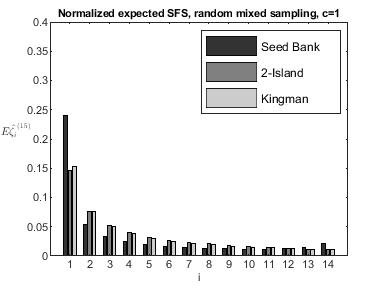}
\includegraphics[width = 0.49\textwidth]{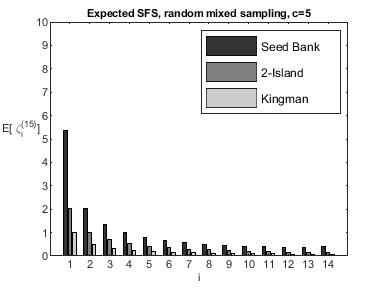}
\includegraphics[width = 0.49\textwidth]{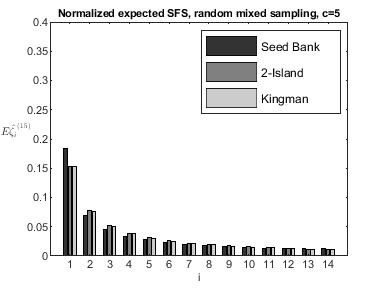}
\end{figure}
\begin{figure}
\centering
\includegraphics[width = 0.49\textwidth]{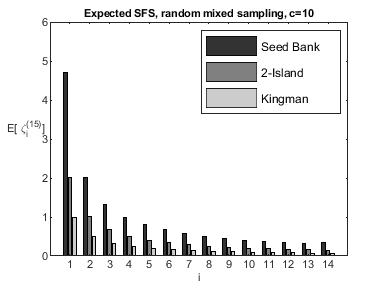}
\includegraphics[width = 0.49\textwidth]{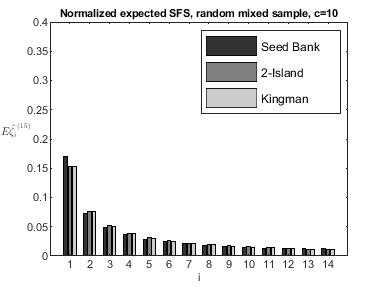}
\caption{Expected SFS sampled from the whole population, i.e.\ $\pi^{\text{unif}}$, with $K = u =1$.}
\label{fig:12}
\end{figure}

\section{Recursions for the sampling distributions}
\label{sampling_distributions}

In this section we use recursions to characterize the (in general intractable) sampling distributions for scenario $\tS$ under the IAM and ISM. 
Similar results under the FAM are provided in the appendix.
The corresponding recursions for $\tK$, $\tW$, and $\tTI$ are special cases of \cite[Eq (2)]{DIG4}.
We will also describe a low-variance Monte Carlo scheme to approximate solutions of these recursions, and hence conduct unbiased inference and model selection based on full likelihoods.

\subsection{IAM recursion}

Let $p(\n^{(1)}; \n^{(2)})$ be the probability of observing sample $\n^{(1)}$ from the active population, and $\n^{(2)}$ from the seed bank under $\tS$, and $\e_i$ be the canonical unit vector with a 1 in the $i$th place, and zeros elsewhere.
Then $p(\n^{(1)};\n^{(2)})$ solves
\begin{align*}
&\left[n^{(1)} \left( \frac{ n^{(1)} - 1 }{ 2 } + u + c \right) + n^{(2)} (u' + K c ) \right] p(\n^{(1)}; \n^{(2)}) \\
&= u n^{(1)} \sum_{i : (n_i^{(1)}, n_i^{(2)}) = (1, 0)} p(\n^{(1)} - \e_i; \n^{(2)}) \\
&\phantom{=} + u' n^{(2)} \sum_{i : (n_i^{(1)}, n_i^{(2)}) = (0, 1)} p(\n^{(1)}; \n^{(2)} - \e_i) \\
&\phantom{=} + \frac{ n^{(1)} }{ 2 } \sum_{i : n_i^{(1)} \geq 2} (n_i^{(1)} - 1) p(\n^{(1)} - \e_i; \n^{(2)}) \\
&\phantom{=} + c n^{(1)} \sum_{i : n_i^{(1)} \geq 1} \frac{n_i^{(2)} + 1}{n^{(2)} + 1} p(\n^{(1)} - \e_i; \n^{(2)} + \e_i) \\
&\phantom{=}+ K c n^{(2)} \sum_{i : n_i^{(2)} \geq 1} \frac{n_i^{(1)} + 1}{n^{(1)} + 1} p(\n^{(1)} + \e_i; \n^{(2)} - \e_i),
\end{align*}
with boundary condition $p(\e_i; 0) = p(0; \e_i) = 1$.
This recursion can be obtained from \cite[Eq (2)]{DIG4} by omitting those transitions which are not allowed in $\tS$, and adjusting the coefficient on the left hand side accordingly.

\subsection{ISM recursion}

The $\tS$ sampling recursion under the ISM is
\begin{align*}
&\left[n^{(1)} \left(\frac{n^{(1)} - 1}{2} + u + c\right) + n^{(2)} (u' + K c) \right]  p(\t, \n^{(1)}, \n^{(2)}) \\
&= u \sum_{\substack{i : n_i^{(1)} = 1, n_i^{(2)} = 0 \\ s_1^{(k)}(t_i) \neq t_j \forall j \forall k}} p(s_i^{(k)}(\t), \n^{(1)}, \n^{(2)}) + u' \sum_{\substack{i : n_i^{(1)} = 0, n_i^{(2)} = 1 \\ s_1^{(k)}(t_i) \neq t_j \forall j \forall k}} p(s_i^{(k)}(\t), \n^{(1)}, \n^{(2)}) \\
&\phantom{=} + u \sum_{i : (n_i^{(1)}, n_i^{(2)}) = (1, 0)} \sum_{(j, k) : s_1^{(k)}(t_i) = t_j} (n_j^{(1)} + 1) p(d_i(\t ), d_i(\n^{(1)} + \e_j), d_i(\n^{(2)})) \\
&\phantom{=} + u' \sum_{i : (n_i^{(1)}, n_i^{(2)}) = (0, 1)} \sum_{(j, k) : s_1^{(k)}(t_i) = t_j} (n_j^{(2)} + 1) p(d_i(\t ), d_i(\n^{(1)}), d_i(\n^{(2)} + \e_j)) \\
&\phantom{=} + n^{(1)} \sum_{i : n_i^{(1)} \geq 2} \frac{n_i^{(1)} - 1}{2} p(\t, \n^{(1)} - \e_i, \n^{(2)}) \\
&\phantom{=} + c n^{(1)} \sum_{i : n_i^{(1)} \geq 1} \frac{n_i^{(2)} + 1}{n^{(2)} + 1} p(\t, \n^{(1)} - \e_i, \n^{(2)} + \e_i) \\
&\phantom{=} + K c n^{(2)} \sum_{i : n_i^{(2)} \geq 1} \frac{n_i^{(1)} + 1}{n^{(1)} + 1} p(\t, \n^{(1)} + \e_i, \n^{(2)} - \e_i),
\end{align*}
with boundary condition $p(\emptyset, (1), (0)) = p(\emptyset, (0), (1)) = 1$, and where $s_i^{(k)}(\t)$ removes the $k^{\text{th}}$ element of $t_i$, e.g.\
\begin{equation*}
s_1^{(2)}((\{0, 2, 3\},\{1\})) = (\{0, 3\},\{1\}),
\end{equation*}
while $d_i(\t)$ removes $t_i$ entirely, e.g.\
\begin{equation*}
d_1((\{0, 2, 3\},\{1\})) = (\{1\}).
\end{equation*}

\subsection{A Monte Carlo scheme for solving sampling recursions}\label{monte_carlo_scheme}

The $\tK$ and $\tW$ coalescents under either IAM or parent-independent FAM are the only instances for which sampling recursions can be solved explicitly.
Numerical schemes for solving the recursions directly also fail for moderate sample sizes because of combinatorial explosion of the number of equations.
Hence, Monte Carlo schemes are used to approximate solutions in practice.
One example of such a scheme is importance sampling, briefly introduced below.

Let $\{\H_g\}_{g = 0}^G$ denote the history of a sample $\n$, so that $\H_0 = \n$, $\H_G$ is the type of the most recent common ancestor, and $\H_{g + 1}$ differs from $\H_g$ by one coalescence, mutation, or migration event.
Then the likelihood of the sample can be written as
\begin{align}
p(\n) &= \sum_{\H_0, \ldots, \H_G} p(\n | \H_0, \ldots, \H_G) \P(\H_0, \ldots, \H_G) \nonumber \\
&= \sum_{\H_0} \ldots \sum_{\H_G} p(\n | \H_0, \ldots, \H_G) p(\H_G) \prod_{g=1}^G \P(\H_{g - 1} | \H_g). \label{generic_recursion}
\end{align}
All of the recursions presented above are of this form, with $p(\n | \H_0, \ldots, \H_G) = \mathds{1}(\H_0 = \n)$, with the coefficients of the recursions denoting the transition probabilities $\P(\H_{g - 1} | \H_g)$, and with $p(\H_G)$ corresponding to the boundary conditions.
A naive Monte Carlo scheme for approximating this sum might sample a most recent common ancestor from the law $p(\H_G)$, evolve the sample stochastically until it reaches the desired size $n + 1$ with probabilities given by the coefficients of the appropriate sampling recursion, and then evaluate the quantity of interest $\mathds{1}(\H_0 = \n)$, where $\H_0$ is the last sample with size $n$.
However, likelihoods in genetics can be vanishingly small, which renders the number of such simulations required for accurate estimators infeasibly large.
Instead, we introduce an importance sampling proposal distribution $\Q(\H_g | \H_{g - 1})$, which acts in the opposite direction of time to $\P(\H_{g - 1} | \H_g)$, i.e.~from the observed leaves towards the most recent common ancestor, and rewrite the summation in \eqref{generic_recursion} as 
\begin{equation*}
p(\n) = \sum_{\H_0} \ldots \sum_{\H_G} p(\H_G) \prod_{g=1}^G \frac{\P(\H_{g - 1} | \H_g)}{\Q(\H_g | \H_{g - 1})} \Q(\H_g | \H_{g - 1}).
\end{equation*}
We will specify $\Q$ in such a way that $\Q(\H_0 = \n) = 1$, which is why the factor $p(\n | \H_0, \ldots, \H_G)$ no longer appears.
This initial condition is then propagated back to the most recent common ancestor with yet-to-be-specified transition probabilities $\Q(\H_g | \H_{g - 1})$, and once the most recent common ancestor is reached, we evaluate the modified quantity of interest 
\begin{equation*}
p(\H_G) \prod_{g = 1}^G \frac{\P(\H_{g - 1} | \H_g)}{\Q(\H_g | \H_{g - 1})}.
\end{equation*}
Every sample results in a positive contribution under this scheme, reducing the variance of estimators.
Careful choices of $\Q$ can reduce variance even further.

The zero-variance proposal distribution $\Q$ under $\tK$ (and thus also $\tW$) was described in \cite{Stephens00}, and extended to $\tTI$ in \cite{DIG4}.
Neither can be implemented, but both articles also provide heuristic approximations from which ancestral coalescence, mutation, and migration events can be sampled, and which result in low variance estimators in practice.
In this section we present similar heuristics for $\tS$ under the IAM and ISM.
As before, corresponding results under the FAM are provided in the appendix.

For the IAM and ISM, we suggest the following procedure for sampling the next event backwards in time given that the current state is $( \n^{ ( 1 ) }, \n^{ ( 2 ) } )$:
\begin{enumerate}
\item Sample the active or dormant subpopulation with probabilities proportional to
\begin{equation*}
\left( n^{ ( 1 ) } \left( \frac{ n^{ ( 1 ) } - 1 }{ 2 } + c + u \right), n^{ ( 2 ) } ( K c + u' ) \right).
\end{equation*}
Denote the chosen subpopulation by $j$.
\item Sample a lineage uniformly at random from subpopulation $j$. Denote its allele by $i$.
\item With probabilities proportional to 
\begin{equation*}
\Bigg( \frac{ ( n_i^{ ( j ) } - 1 )^+ }{ 2 } \mathds{ 1 }_{ \{ j = 1 \} }, u \mathds{ 1 }_{ \{ j = 1 \} } + u' \mathds{ 1 }_{ \{ j = 2 \} }, c \mathds{ 1 }_{ \{ j = 1 \} } + K c \mathds{ 1 }\{ j = 2 \} \Bigg),
\end{equation*}
merge the lineage with another one with allele $i$ on island $j$, remove from type $i$ a randomly chosen mutation that does not appear on any other lineage, or migrate the lineage to the other subpopulation.
The mutation probability is taken to be 0 if there are no eligible mutations on the lineage, or if the frequency of the allele is greater than one in the case of the IAM.
For the IAM, we also interpret the removal of a mutation as the removal of the lineage from the sample.
\end{enumerate}

\section{Inference and model selection}
\label{model_selection}

In this section we provide an example of the impact of the presence or absence of a seed bank on model selection, and on estimating coalescent parameters from genetic data.
Our focus is on the full likelihood Monte Carlo methods introduced in Section \ref{monte_carlo_scheme}, rather than on summary statistics such as the (n)SFS.
While computationally intensive, this choice lets us draw robust conclusions about the extent to which DNA sequence data can distinguish between our three model classes even in principle, without further confounding by the limitations of any particular summary statistic.

\subsection{Model selection based on sampling formulas}
\label{ssn:Exact_SFS_likelihood}

We used a pseudo-marginal Metropolis-Hastings algorithm \cite{AR09} to perform model selection and parameter inference simultaneously for models $\tK$, $\tS$, and $\tTI$ using the full likelihood of the observed sequence data.
Model $\tW$ was not included as it is not identifiable from $\tK$.
We focus on the ISM in order to balance biological relevance and computational cost.
Data set of 100 observed, non-recombining sequences with were simulated under each model and various parameter regimes to act as observed data.
In each case, all 100 sequences were sampled from island 1 to model the impact of an unknown seed bank or population subdivision.

The state space of our pseudo-marginal Markov chain consists of the model indicator $\tI \in \{ \mathtt{K}, \mathtt{S}, \mathtt{TI}\}$, as well as seven non-negative variables
\begin{equation*}
\Theta := ( u_{\tK}, u_{\tS}, u_{\tTI}, c_{\tS}, c_{\tTI}, K_{\tS}, K_{\tTI} ).
\end{equation*}
In particular, the fact that $u' = 0$ under $\tS$ and $\tTI$ was assumed to be known.
Given an observed data set $( \t, \n )$, the target distribution is the posterior
\begin{equation*}
q( \tI, \Theta | \t, \n ) \propto p( \t, \n | \tI, \Theta ) q_{\tI}( \tI ) q_{ u_{\tK}}( u_{\tK} ) \prod_{ J \in \{ \tS, \tTI \} } q_{ u_J }( u_J ) q_{ c_J }( c_J ) q_{ K_J }( K_J ),
\end{equation*}
where $\mathbf{ n } = ( \mathbf{ n }^{ ( 1 ) }, \mathbf{ n }^{ ( 2 ) } )$ in the case of scenarios \texttt{S} and \texttt{TI}.
Here, the likelihood $p( \t , \n | \tI, \Theta )$ only charges those coordinates of $\Theta$ that play a role for model $\tI$, and is flat in all other directions.
The prior distributions are $q_{\tI} = ( 1 / 3, 1 / 3, 1 / 3 )$, and Gamma-distributions with shape parameter 4 for all other variables. 
Scale parameters are fixed at $1/4$ for the $c$ and $K$-variables, and by requiring the prior mean to equal the corresponding Watterson estimator for the $u$-variables.
This updating of locally redundant variables increases model dimension, but also results in faster mixing across the three different models since all parameters are updated simultaneously (see the ``saturated space approach'' of \cite{BGR03}), and accounts for the fact that the same number of segregating sites can fit two very different mutation rates in different model classes.

The model index was resampled uniformly at random at each time step, including the possibility of remaining in place.
All other parameters were updated using independent Gaussian increments with mean 0 and variance $\approx  1 / 14$, with all parameters reflected at zero.
The importance sampling scheme of Section \ref{monte_carlo_scheme} was used to obtain unbiased estimates of likelihoods, with particle numbers set to 400 for $\tK$, and 20 000 for $\tS$ and $\tTI$.
Variances of estimators were further reduced by employing stopping time resampling \cite{J12}.
These parameters were calibrated so that the log-likelihood estimator variances were close to 3, and acceptance probabilities close to 7\%, shown to be optimal in \cite{STRR15}.
\texttt{C++} code for both simulating observed data sets, and conducting the inference described above, is available at \url{https://github.com/JereKoskela/seedbank-infer}.

Three realizations of this Markov chain, one for each simulated data set, were run for 100 000 steps each, initialized from a uniformly chosen model, and the continuous parameters initialized from their respective prior means.
The most immediate question is whether each data-generating model can be correctly recovered from its observed data set.
Table \ref{marginal_posteriors} provides marginal posterior probabilities of each model and data set.
It is evident that the true model can be recovered from a moderate amount of data with high confidence for moderate second population sizes and migration rates.
However, as might be expected, large migration rate or small second populations make it challenging to tell the three models apart, at least when only sampling a single locus from one population.
\begin{table}[h!]
\centering
\begin{tabular}{ c | c c c | c c c }
True model & $q_{\tI}(\tK| \mathbf{ t }, \mathbf{ n })$ & $q_{\tI}(\tS| \mathbf{ t }, \mathbf{ n })$ & $q_{\tI}(\tTI | \mathbf{ t }, \mathbf{ n } )$ & $u_0$ & $c_0$ & $K_0$ \\
\hline
\texttt{K} & 0.950 & 0.042 & 0.008 & 10 & - & - \\
\texttt{S} & 0.000 & 1.000 & 0.000 & 10 & 1 & 1 \\
\texttt{TI} & 0.132 & 0.027 & 0.841 & 10 & 1 & 1 \\
\texttt{S} & 0.258 & 0.463 & 0.279 & 10 & 5 & 1 \\
\texttt{TI} & 0.439 & 0.028 & 0.533 & 10 & 5 & 1 \\
\texttt{S} & 0.224 & 0.519 & 0.257 & 10 & 1 & 5 \\
\texttt{TI} & 0.356 & 0.137 & 0.507 & 10 & 1 & 5 \\
\end{tabular}
\caption{Marginal posterior probabilities of each model class for given data-generating parameters.}
\label{marginal_posteriors}
\end{table}

Posterior distributions of parameters given a model class are also of interest.
These are summarized in Table \ref{parameter_posteriors}.
None of the parameters are strongly identified, and in some cases estimates are biased due to the fact that our data set consists of only a single locus.
The mutation rate was the slowest to mix in all cases (results not shown).

\begin{table}[h!]
\centering
\begin{tabular}{ c | c c c c | c c c c  }
  & 0.025 & 0.5 & 0.975 & Model & 0.025 & 0.5 & 0.975 & Model \\
  \hline
$u$ & 6.30 & 8.97 & 11.3 & \tK \\
$u$ & 7.45 & 8.60 & 9.87 & \tS   & 13.0 & 16.3 & 19.7 & \tTI \\
$c$ & 0.20 & 0.98 & 1.86 & \tS & 0.39 & 1.02 & 1.94 & \tTI \\
$K$ & 0.09 & 0.54 & 1.72 & \tS & 0.25 & 0.84 & 1.82 & \tTI  \\
$u$ & 8.32 & 8.59 & 9.06 & \tS, $c = 5$ & 6.26 & 9.69 & 12.9 & \tTI, $c = 5$\\
$c$ & 2.00 & 3.44 & 4.23 & \tS, $c = 5$ & 2.04 & 5.06 & 9.02 & \tTI, $c = 5$\\
$K$ & 0.57 & 1.04 & 1.37 & \tS, $c = 5$ & 0.35 & 0.86 & 1.87 & \tTI, $c = 5$ \\
$u$ & 4.06 & 6.03 & 8.97 & \tS, $K = 5$ & 8.29 & 13.1 & 16.8 & \tTI, $K = 5$\\
$c$ & 0.36& 1.03 & 2.52 & \tS, $K = 5$ & 0.28 & 1.04 & 2.39 & \tTI, $K = 5$\\
$K$ & 1.69 & 4.23 & 6.91 & \tS, $K = 5$ & 1.68 & 3.47 & 7.01 & \tTI, $K = 5$
\end{tabular}
\caption{Posterior quantiles for various parameters and scenarios. Where not specified, the parameters are $u = 10$, $c = K = 1$. All estimates are conditional on the true model class.}
\label{parameter_posteriors}
\end{table}

\begin{figure}
\centering
\includegraphics[width = 0.49\textwidth]{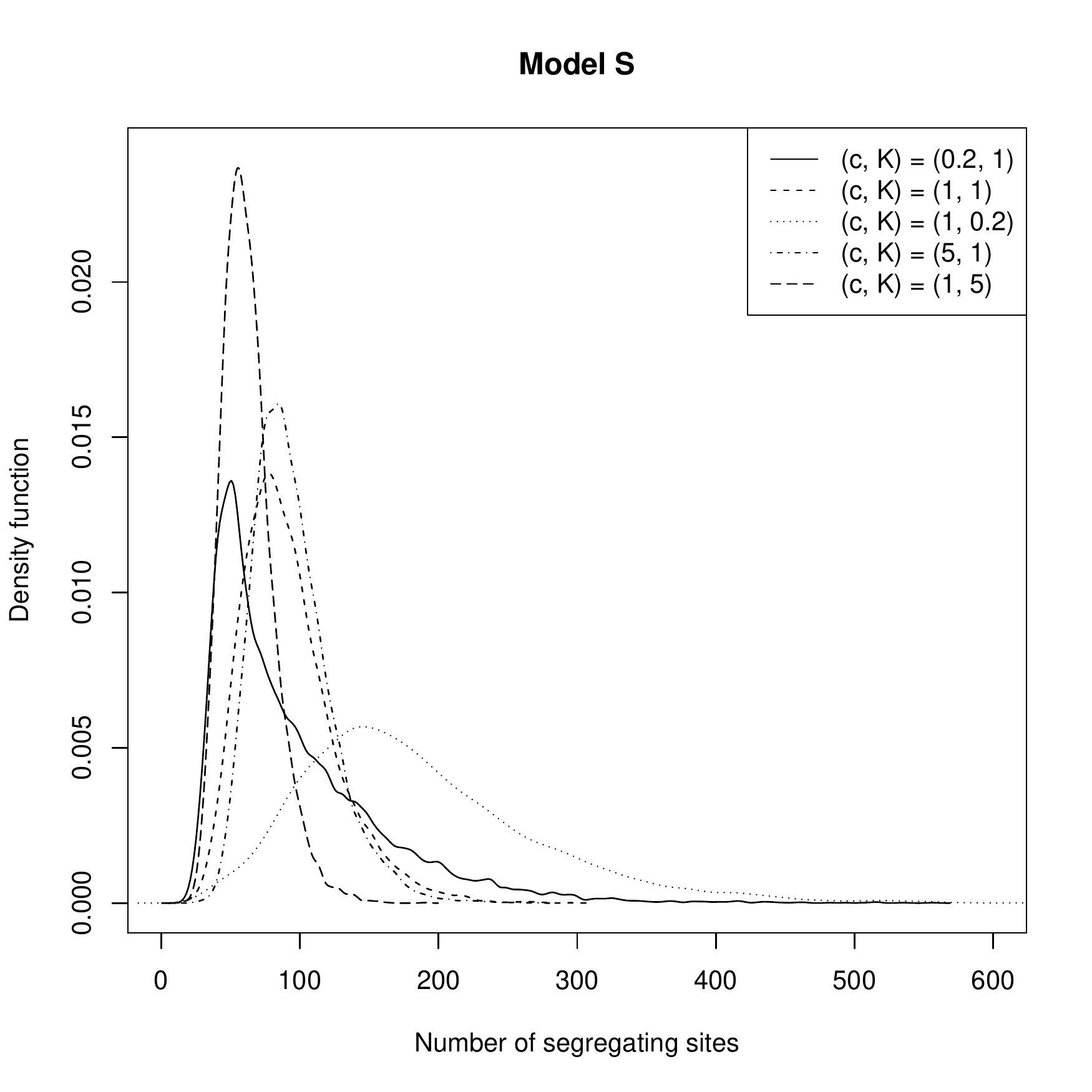}
\includegraphics[width = 0.49\textwidth]{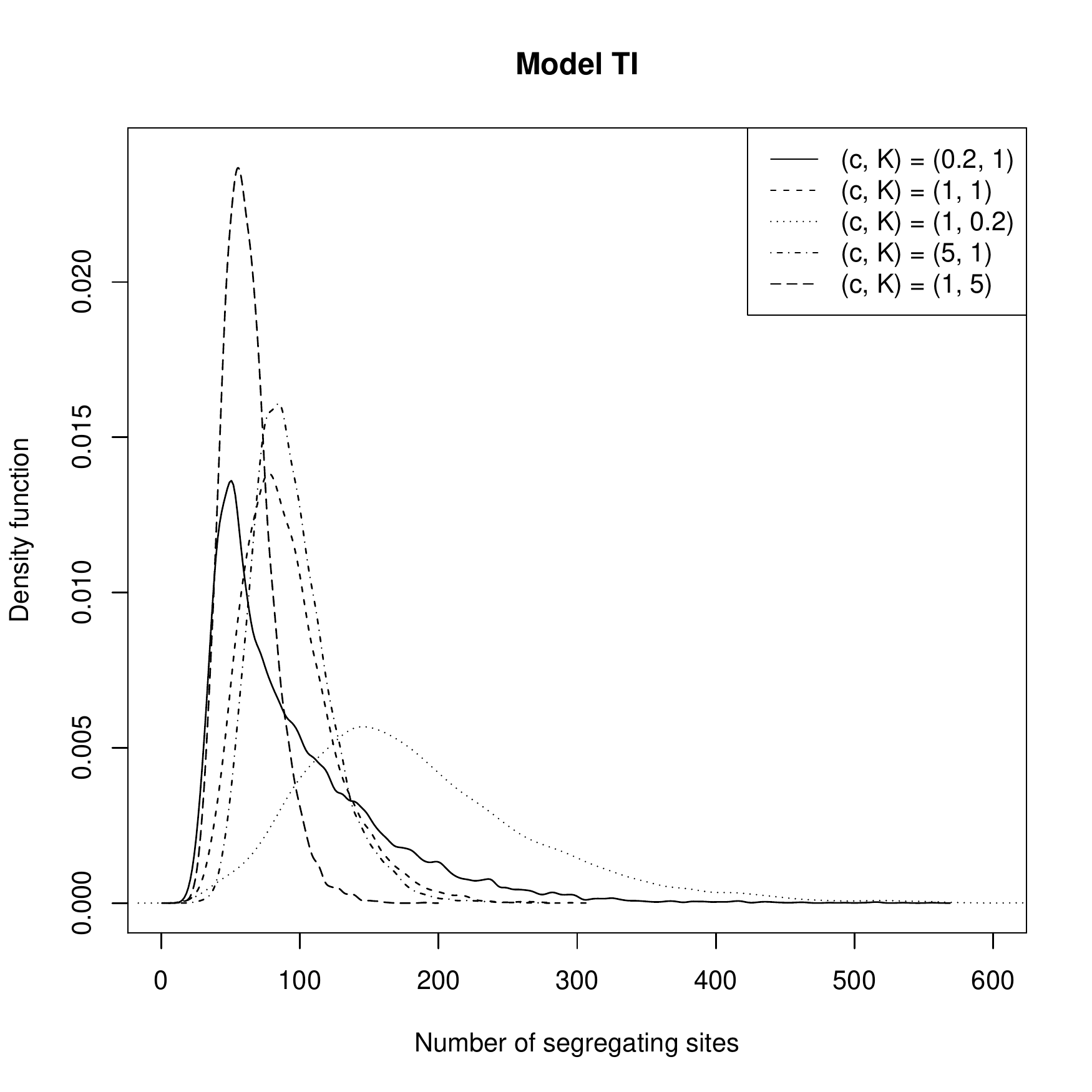}
\caption{Empirical distributions of the number of segregating site among 100 samples from population 1, estimated from 10 000 simulations.}
\label{fig:13}
\end{figure}

Low migration rates and large second populations are also problematic, albeit in a different way.
Figure \ref{fig:13} shows the empirical distribution of the number of segregating sites for samples drawn from models {\tS} and {\tTI}.
Results with $c = 0.2$ or $K = 0.2$ have noticeably broader supports and heavier tails than any of the other scenarios, because a migration from one population to another is rare, but will result in a very long ancestral tree if it takes place.
The consequence for inference is that realisations of data sets are not informative of the model or parameters which generated them, and the importance sampling schemes from Section \ref{monte_carlo_scheme} will also suffer from high variance and very long run times.
This is reminiscent of Figures \ref{fig:10} and \ref{fig:12}, which show that the expected SFS detects an excess of singletons due to a strong seed bank under uniform sampling, but not when only the active population is sampled.

While the method presented in this section does not scale to large data sets, it sets a benchmark for what can be expected of the performance of more scalable methods.
In particular, the three model classes ought to be distinguishable with high confidence even in the presence of parameter uncertainty, provided that the true migration rates and seed bank sizes are moderate.
True values of $c$ or $K$ that are too large are problematic due to less statistical separation between the models, while low data-generating values of $c$ and $K$ cause instability both in observed data and in our Monte Carlo scheme.
Estimating precise values of parameters within model classes is challenging without strong prior information, or data from multiple unlinked loci, neither of which has been used in our model selection pipeline.

\subsection{Detecting mutation in the seed bank}

In this section we focus on a different model selection problem: whether mutation is taking place in a strong seed bank that is known to be present.
Data sets were simulated under two scenarios with a moderate seed bank and migration rate $K = c = 1$:
\begin{enumerate}
\item[S1.] Model $\tS$ with $u = 10, u' = 0$.
\item[S2.] Model $\tS$ with $u = u' = 5$.
\end{enumerate}
All other simulation details are as in Section \ref{ssn:Exact_SFS_likelihood}.
A pseudo-marginal Metropolis-Hastings chain was run targeting these two hypotheses, with the same priors as in Section \ref{ssn:Exact_SFS_likelihood}.
In scenario S1 we assumed that $u' = 0$ was known, while in scenario S2 we assumed that $u = u'$ was known, but that the common value itself was not.
The posterior probabilities of each scenario are given in Table \ref{marginal_posterior_mutation}.
\begin{table}[h!]
\centering
\begin{tabular}{ c | c c }
True scenario & $q_I(S1| \mathbf{ t }, \mathbf{ n })$ & $q_I(S2| \mathbf{ t }, \mathbf{ n })$ \\
\hline
\texttt{S1} & 1.000 & 0.000 \\
\texttt{S2} & 0.098 & 0.902 
\end{tabular}
\caption{Marginal posterior probabilities of each scenario.}
\label{marginal_posterior_mutation}
\end{table}

It is evident that the presence or absence of mutation in a seed bank can be detected with high confidence from a modest amount of data.
Table \ref{mutation_parameter_posteriors} shows that parameters remain only weakly identified and that estimates can be biased, particularly in the case of mutation rates.
Once again, mutation rates were also the slowest parameters to mix.

\begin{table}[h!]
\centering
\begin{tabular}{ c | c c c c  }
  & 0.025 & 0.5 & 0.975 & Scenario \\
\hline
$u $ & 3.42 & 5.82 & 9.02 & S1 \\
$c $ & 0.17 & 0.98 & 2.89 & S1  \\
$K$ & 0.15 & 0.80 & 1.98 &  S1 \\
$u = u'$ & 2.29 & 5.54 & 8.57 & S2\\
$c $ & 0.24 & 1.04 & 2.23 &  S2 \\
$K $ & 0.28 & 0.92 & 2.04 & S2
\end{tabular}
\caption{Posterior quantiles for various parameters and scenarios. The columns labeled $( u_0, c_0, K_0 )$ denote the corresponding data-generating parameters.}
\label{mutation_parameter_posteriors}
\end{table}

\section{Discussion}
\label{sec:discussion}

We have reviewed several population genetic models related to seed banks, in combination with several classical mutation models.
We derived expressions for classical population genetic summary statistics such as the $F_{ST}$ and the SFS for various combinations of coalescent and mutation models.
We then studied the identifiability of various scenarios and parameters based on tractable summary statistics, as well as computationally intensive full likelihood methods.
Throughout, our focus is on deriving and testing generic methodology without prior assumptions on mutation models, parameter ranges, or model classes.
This is to facilitate analysis of sequence data across a wide range of scenarios and species.

Explicit expressions for $F_{ST}$ for the IAM and ISM can be obtained using phase-type distribution arguments \cite{HSJB18}.
A strong seed bank produces elevated levels of $F_{ST}$, but less so than the two-island model with identical parameters. 
The signal is slightly stronger in the case without mutation in the seed bank compared to the case with mutation, but generally appears to be too weak to allow for confident detection of a strong seed bank. 
The FAM also yields similar results (see appendix).

Considering the normalized SFS instead of $F_{ST}$ results in improved statistical power.
The Kingman and the weak seed bank scenarios can only be distinguished with prior knowledge of the germination rate or population-rescaled mutation rate(s), whereupon the number of expected segregating sites suffices as a statistic. 
The strong seed bank and two island models result in an excess of singletons and a lighter tail in the nSFS when compared to the classical Kingman case, for sample sizes as low as $n = 15$. 
Thus, these two scenarios can be distinguished from $\tK$ and $\tW$, but not from each other. 
The deviation of the nSFS from the Kingman coalescent arises due to the reduced (or vanishing) mutation rate in the seed bank, and so sampling dormant lineages is an effective way to boost the power of the model selection procedure.

To study the scope of possible inference, we used a Monte Carlo scheme to approximate likelihoods of observed sequence data.
For moderate data-generating parameter values, model selection from simulated data gave good results for samples of size $n = 100$ and a single locus, even in the presence of parameter uncertainty. 
Accounting for parameter uncertainty in the simulation pipeline is particularly important, because standard estimators such as the Watterson estimator assume a fixed coalescent model, and thus using the wrong estimator can strongly bias further inferences.
Frequent migration or a small relative second sub-population size cause diminished statistical power, while rare migrations or a large second sub-population cause instabilities in both the model and in standard importance sampling estimators of likelihoods.

We have also demonstrated that our method is able to detect whether mutation is taking place in the seed bank, again in the presence of parameter uncertainty.
Thus, it provides a promising first step towards answering similar questions in general \cite{LJ11}, such as assessing molecular clock hypotheses for bacteria, or other organisms without easy access to a fossil record \cite{M07}.

Knowledge of the real substitution rate $\hat \mu$ per year at the (active) locus under consideration would allow a real-time embedding of the coalescent history via
$$
\mbox{coalescent time unit } \times  u_{\tI} \approx \mbox{ year } \times \mu, 
$$
for $\tI \in \{ \tK, \tW, \tS, \tTI \}$ (see e.g.~\cite[Eq (4)]{EBBF15}, \cite[Section 4.2]{SBB13}).
This allows the estimation of quantities such as the TMRCA of a sample in real time, not only in units of coalescent time.
Typically, one coalescent time unit corresponds to $O(N)$ generations under all four models considered in this paper.

Our paper is a starting point for the statistical methodology for seed bank detection. 
We have shown that model selection and inference are possible from moderate data sets in principle, but several important points remain to be addressed.

First, the adequacy and universality of the models needs to be established. 
They all describe idealized scenarios in population genetics, with constant population sizes, and in the absence of further evolutionary forces such as selection or demography. 
The effect of such forces in the presence of seed banks remains unknown, and may confound some or all of the results we have presented.
Indeed \cite{ZT12, SAMT19} have shown that weak seed banks and demographic changes confound each other unless considered jointly, and similar results are available for the structured coalescent \cite{MRGBC16, RMGACBC18}.
A similar analysis for the strong seed bank model, and the effect of a misspecified model on demographic inference, remains an open problem. A layer of complexity is added in the modeling of demography with a seed bank through the effect of changes in demography for the dormant population. In order to obtain a (time-changed) coalescent with a strong seed bank mechanism, the demographic changes would have to equally affect the active and the dormant population. If, however, the seed bank remains constant relative to the demographic changes in the active population, the result would be a coalescent with a time-inhomogeneous (de)activation mechanism.

Second, the type of seed bank formation mechanism itself needs to be discussed. 
The strong seed bank model of \cite{BGCKW16} analyzed here follows the modeling idea of ``spontaneous switching'' in \cite{LJ11}, where switching between the active and the dormant state happens on an individuals basis. 
\cite{LJ11} argue that this model might be appropriate for populations in ``stable'' environments, but that in real populations initiation of or resuscitation from dormancy can be triggered by environmental cues, leading to ``responsive switching'' where many individuals switch their state simultaneously. This mechanism can be incorporated at the same scale as the spontaneous switching and  leads to a scaling limit that is different from the migration-type behavior of the strong seed bank model (and of course also differs from the weak seed bank model), cf. \cite{BGCKW19}, as it includes simultaneous activation and deactivation of lineages. The effect of this additional mechanism on the statistics discussed here remains to be studied.
%
%

\section*{Acknowledgements}

JB was supported by  DFG Priority Programme 1590 ``Probabilistic Structures in Evolution'', project 1105/5-1.
EB was supported by DFG RTG 1845 and BMS Berlin Mathematical School.
JK was supported in part by EPSRC grant EP/R044732/1.

\section*{Appendix}

\subsection*{Classical measures of population structure under the FAM}

The \emph{sample heterozygosity} $H$ of a population is defined as the probability of two individuals drawn independently and uniformly from the population carrying different alleles. 
For $\tK$ and $\tW$, the stationary sample heterozygosity is
\begin{align*}
H^{ \tK } &:= 2 \E^{ \tK  }[ X ( 1 - X ) ], \qquad \text{and} \qquad H^{ \tW } := 2 \E^{ \tW  }[ X ( 1 - X ) ],
\end{align*}
where $X$ has the stationary distribution of \eqref{eq:system_general} corresponding to each model. 

A well-known result (e.g.\ \cite[p.\ 49]{Et11}) states that 
\begin{align*}
H^{\tK} = \frac{4u_1 u_2}{(u_1+u_2)(1+2u_1+2u_2)},
\end{align*}
an similarly we have the intuitive result
\begin{align*}
H^{\tW}=\frac{4u_1 u_2}{(u_1+u_2)(\beta^2+2u_1+2u_2)}.
\end{align*}

For structured populations one distinguishes between the \emph{global} and \emph{local} sample heterozygosities, corresponding to samples taken from the overall population, resp.~from each sub-population. 
Thus, with $(X,Y)$ being the solution to \eqref{eq:system_general} at stationarity, the local sample heterozygosities for each sub-population under $\tS$ and $\tTI$ are
\begin{align*}
H_X^{ \tS } &:= 2 \E^{ \tS } [ X ( 1 - X ) ], \qquad \qquad & H_X^{ \tTI } &:= 2 \E^{ \tTI } [ X ( 1 - X ) ], \\
H_Y^{ \tS } &:= 2 \E^{ \tS } [ Y ( 1 - Y ) ], \qquad \qquad & H_Y^{ \tTI } &:= 2 \E^{ \tTI } [ Y ( 1 - Y ) ],
\end{align*}
and therefore the global sample heterozygosities can be written as
\begin{align}\label{eq:H_def}
H^{ \tS } := {}& \frac{ K^2 }{ ( K + 1 )^2 }H_X^{ \tS } + \frac{ 2 K }{ ( K + 1 )^2 } \E_{ \mu^{ \tS } }[ X ( 1 - Y ) + Y ( 1 - X ) ] + \frac{ 1 }{ ( K + 1 )^2 }H_Y^{ \tS },\notag \\
H^{ \tTI } := {}& \frac{ K^2 }{ ( K + 1 )^2 }H_X^{ \tTI } + \frac{ 2 K }{ ( K + 1 )^2 } \E_{ \mu^{ \tTI } }[ X ( 1 - Y ) + Y ( 1 - X ) ] \notag\\
&{}+ \frac{ 1 }{ ( K + 1 )^2 } H_Y^{ \tTI },
\end{align}
where the weights on the local heterozygosities are the probabilities associated with sampling two lineages uniformly at random from the global population.
The sample heterozygosity at stationarity is well-studied under the FAM and either $\tK$ or $\tTI$ \cite{H94}, it has so far not been considered for seed banks. 

Note that we can rewrite the sample heterozygosities for $\tI \in \{\tS,\tTI\}$ in terms of mixed moments using the notation
\begin{align*}
M_{n,m}^{\tI}:=\E_{ \mu^{\tI} }[X^n Y^m], \quad n,m \ge 0.
\end{align*}
This immediately gives 
\begin{align*}
H_X^{\tI} = 2(M^{\tI}_{1,0}-M^{\tI}_{2,0}), \qquad\qquad
H_Y^{\tI} = 2(M^{\tI}_{0,1}-M^{\tI}_{0,2}),
\end{align*}
and therefore 
\begin{align*}
H^{\tI} = \frac{2}{(K+1)^2} \Big( (K^2 + K) M^{\tI}_{1,0} + (K+1) M^{\tI}_{0,1} - 2K M^{\tI}_{1,1} - K^2 M^{\tI}_{2,0} - M^{\tI}_{0,2} \Big).
\end{align*}
These mixed moments can be calculated recursively \cite[Lemma 2.7]{BBGKWB19}. 
For example, $M^{\tI}_{0,0} = 1$ and
\begin{align*}
M^{\tI}_{1,0} &= \frac{cu_2'+u_1u_2'+u_2u_2'+cKu_2}{cu_1'+cu_2'+u_1u_1'+u_1u_2'+u_2u_1'+u_2u_2'+cKu_1+cKu_2}, \\
M^{\tI}_{0,1} &= \frac{cu_2'+u_1'u_2+u_2u_2'+cKu_2}{cu_1'+cu_2'+u_1u_1'+u_1u_2'+u_2u_1'+u_2u_2'+cKu_1+cKu_2},
\end{align*}
for the first moments.
These first moments do not depend on $\alpha$ and $\alpha'$, which is clear intuitively since they represent variance parameters. 
Hence, $M^{\tI}_{1, 0}$ and $M^{\tI}_{0, 1}$ are invariant for $\tI \in \{ \tTI, \tS \}$. 
The expression for the second moments can also be computed easily, but are cumbersome and therefore omitted.

In the case of equal relative population sizes ($K=1$), migration rate $c=1$ and mutation rates $u_1=u_2=u_1'=u_2'= 1/2$,  we obtain 
\begin{align*}
H^{\tS}  = \; \frac{14}{31} \approx 0.4516  \; > H^{\tTI} = \frac{13}{32} \approx 0.4063 \;>\;\; \frac{1}{3} =  H^{\tK}. 
\end{align*}
Moreover, using simple sign arguments, we find that these relationships also hold in a more general context: if $u_1=u_1'$, $u_2=u_2'$, and $K=1$, then for all $u_1, u_2, c \geq 0$ we have $H^{\tS} \geq H^{\tTI} \geq H^{\tK}$.
However, in all other cases (e.g.\ $c=u_1=u_2=u_1'=u_2'=1$, $K=0.01$), the second inequality does not hold.
It is also interesting to note that if $\beta^2 < 3 / 14$, then $H^{ \tW } = 1 / ( \beta^2 + 2 ) > H^{ \tS }$, showing that a weak seed bank can generate more heterozygosity than a strong seed bank in some cases.

Overall, scenario $\tS$ has elevated levels of genetic variability relative to $\tTI$ or $\tK$ at stationarity.
The $\tTI$ sample heterozygosity is somewhat lower, which is consistent with the idea that genetic drift in the second island reduces variability.

\begin{rem}
If we naively let $K \to \infty$ (i.e.\ the relative second island size $\to 0$) in equation \ref{eq:H_def}, ignoring the intrinsic dependence of the variables $X$ and $Y$ on this parameter, we recover the sample heterozygosity of $\tK$,
\begin{align*}
H_X^{\tS} &\to H^{\tK}\qquad \text{and} \qquad H_X^{\tTI} \to H^{\tK}.
\end{align*}
This convergence holds in a stronger sense on the diffusion level, and will be discussed theoretically in related future work. 
\end{rem}

\begin{rem}
The stationary sample heterozygosity cannot distinguish between $\tK$ and $\tW$ if neither the germination rate $\beta$ nor the population-rescaled mutation rate $u$ are known. 
But $\tK$ and $\tW$ can be differentiated using, for example, the {\em rate of decay} of sample heterozygosity over time \emph{in the absence of mutation}. 
Define
\begin{equation*}
H^{\tI}(t, x) := 2 \E^{\tI}[ X(t) ( 1 - X( t ) )| X(0)=x ],
\end{equation*}
for $I \in \{ \tK, \tW \}$.
Then we obtain
\begin{align*}
H^{\tK}(t, x) &= 2e^{-t} x (1-x), \quad \mbox{ while } \quad  H^{\tW}(t, x) = 2e^{-\beta^2 t} x (1-x),
\end{align*}
showing that $H^{ \tW }$ decays more slowly than $H^{ \tK }$ due to the seed bank slowing down genetic drift \cite{KKL01}.
\end{rem}

\paragraph{Wright's $F_{ST}$ for the FAM}
In the previous section we derived the sample heterozygosities, i.e.\ the probabilities of sampling \emph{distinct} types, in the FAM. 
The probabilities of sampling \emph{identical} types are simply their complements, yielding
\begin{align*}
F_{ST}^{\tI} = \frac{ ( K + 1 ) H^{\tI} - KH_X^{\tI} - H_Y^{ I } }{ ( K + 1 ) H^{\tI} }
\end{align*}
for $\tI \in \{ \tS, \tTI \}$.
For example, fixing $u_1=u_2=1/2=u'_1=u'_2, c=K=1$ and $\alpha =1$, $\tTI$ $(\alpha'=1)$ leads to a stronger differentiation than $\tS$  $(\alpha'=0)$, 
\begin{equation*}
F_{ST}^{ \tS }= \frac{1}{28} < \frac{1}{13} = F_{ ST }^{ \tTI },
\end{equation*}
again indicating that strong seed banks introduce some population substructure, but that the effect is stronger in the two island model. 
This is intuitive, because the dynamics of the population are closer to those of two independent subpopulations when both demes undergo genetic drift than when only one subpopulation does.

\begin{figure}[!ht]
\centering
\includegraphics[width = 0.49\textwidth]{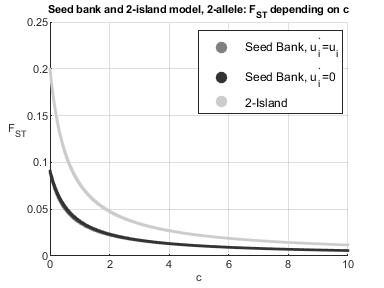}
\includegraphics[width = 0.49\textwidth]{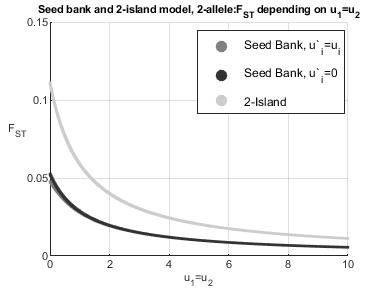}
\includegraphics[width = 0.49\textwidth]{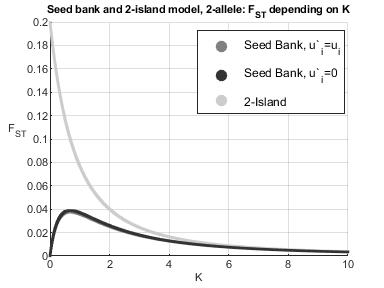}
\caption{ $F_{ST}$ under $\tS$ and $\tTI$ as a function of various parameters in the FAM. Where not specified, $K=c=1$, $u_1=u_2=0.5$.}
 \label{fig:1}
\end{figure}

Figure \ref{fig:1} further illustrates how $F_{ST}$ depends on the model parameters in both cases.
The first plot shows $F_{ST}$ as a function of the migration rate $c$.
As expected, $F_{ST}$ approaches 0 as $c$ increases, leading to a well-mixed population, and the $F_{ST}$ of $\tTI$ dominates the one of $\tS$ by a factor of approximately 2.1 for these parameters.
The second plot shows $F_{ST}$ as a function of the mutation rate, with similar results.
This is again in accordance with expectation, since increasing mutation rates in both subpopulations further mixes the population.
The third plot shows the dependence of $F_{ST}$ on the relative population size $K$. 
The $F_{ST}$ is nearly 0 if the relative population size on either island is very small (i.e.\ $K$ very small or very large), as this results in a small probability of sampling two individuals from different demes when sampling uniformly from the whole population. 

In the absence of mutation in the seed bank, $u' = 0$, and with the parameters $u_1=u_2=1/2, K=c=1$, we get 
\begin{align*}
F_{ST}^{\tS} = \frac{1}{27} > \frac{1}{28},
\end{align*}
a {\em slightly} stronger signal than in the case with mutation. 
The relationship between $K$, $c$ and the $F_{ST}$ in this setting is also illustrated in Figure \ref{fig:1}.

\subsection*{Recursions for the FAM sampling distribution}

Under $\tS$ and the FAM, the sampling distribution $p(\n^{(1)}; \n^{(2)})$ solves 
\begin{align*}
&\left[n^{(1)} \left(\frac{ n^{(1)} - 1 }{ 2 } + u_1 + u_2 + c\right) + n^{(2)} (u_1' + u_2' + K c )\right] p(\n^{(1)}; \n^{(2)}) \\
&= u_2 (n_1^{(1)} + 1) \mathds{1}(n_2^{(1)} > 0) p(\n^{(1)} + \e_1 - \e_2; \n^{(2)}) \\
&\phantom{=} + u_1 (n_2^{(1)} + 1) \mathds{1}(n_1^{(1)} > 0) p(\n^{(1)} - \e_1 + \e_2; \n^{(2)}) \\
&\phantom{=} + u_2' (n_1^{(2)} + 1) \mathds{1}(n_2^{(2)} > 0) p(\n^{(1)}; \n^{(2)} + \e_1 - \e_2) \\
&\phantom{=} + u_1' (n_2^{(2)} + 1) \mathds{1}(n_1^{(2)} > 0) p(\n^{(1)}; \n^{(2)} - \e_1 + \e_2) \\
&\phantom{=} + n^{(1)} \frac{n_1^{(1)} - 1}{2} p(\n^{(1)} - \e_1; \n^{(2)}) + n^{(1)} \frac{n_2^{(1)} - 1}{2} p(\n^{(1)} - \e_2; \n^{(2)}) \\
&\phantom{=} + c n^{(1)} \frac{n_1^{(2)} + 1}{n^{(2)} + 1} \mathds{1}(n_1^{(1)} > 0) p(\n^{(1)} - \e_1; \n^{(2)} + \e_1) \\
&\phantom{=} + c n^{(1)} \frac{n_2^{(2)} + 1}{n^{(2)} + 1} \mathds{1}(n_2^{(1)} > 0) p(\n^{(1)} - \e_2; \n^{(2)} + \e_2) \\
&\phantom{=} + K c n^{(2)} \frac{n_1^{(1)} + 1}{n^{(1)} + 1} \mathds{1}(n_1^{(2)} > 0) p(\n^{(1)} + \e_1; \n^{(2)} - \e_1) \\
&\phantom{=} + K c n^{(2)}  \frac{n_2^{(1)} + 1}{n^{(1)} + 1} \mathds{1}(n_2^{(2)} > 0) p(\n^{(1)} + \e_2; \n^{(2)} - \e_2),
\end{align*}
where $\mathds{1}(E) = 1$ if event $E$ is true, and 0 otherwise.
Boundary conditions are typically prescribed as the stationary distribution specified by the mutation rates, at least when $u_1 = u_1'$ and $u_2 = u_2'$:
\begin{align*}
p((1, 0); (0, 0)) &= p((0, 0); (1, 0)) = \rho_1, \\
p((0, 1); (0, 0)) &= p((0, 0); (0, 1)) = \rho_2.
\end{align*}

\subsection*{A Monte Carlo scheme for the FAM recursions}

Let $p_i(\e_j | \n^{(1)}, \n^{(2)})$ denote the probability that a further lineage sampled from island $i \in \{1, 2\}$ carries allele $j \in \{1, 2\}$, given observed allele frequencies $\n^{(1)}, \n^{(2)}$ from islands 1 and 2, respectively.
These conditional sampling distributions are intractable, but as outlined in Section \ref{monte_carlo_scheme}, approximating them will produce efficient algorithms.

Let
\begin{equation*}
D(n^{(1)}, n^{(2)}) := n^{(1)} \left( \frac{ n^{(1)} - 1 }{ 2 } + u + c \right) + n^{(2)} ( u' + K c ).
\end{equation*}
A calculation similar to \cite[Theorem 1]{Stephens00} identifies the zero-variance proposal distribution for the FAM as
\begin{align*}
(\n^{(1)}, \n^{(2)}) \mapsto (\n^{(1)} - \e_i, \n^{(2)}) &\text{ w.~prob.~} \frac{n_i^{(1)} (n_i^{(1)} - 1) / 2}{p_1( \mathbf{e}_i | \n^{(1)} - \e_i, \n^{(2)}) D(n^{(1)}, n^{(2)})}, \\
(\n^{(1)}, \n^{(2)}) \mapsto (\n^{(1)} - \e_i + \e_j, \n^{(2)}) &\text{ w.~prob.~} \frac{u n_i^{(1)} p_1(\e_j | \n^{(1)} - \mathbf{e}_i, \n^{(2)})}{p_1(\e_i | \n^{(1)} - \e_i, \n^{(2)}) D(n^{(1)}, n^{(2)})}, \\
(\n^{(1)}, \n^{(2)}) \mapsto (\n^{(1)}, \n^{(2)} - \e_i + \e_j) &\text{ w.~prob.~} \frac{u' n_i^{(2)} p_2(\e_j | \n^{(1)}, \n^{(2)} - \mathbf{e}_i)}{p_2(\e_i | \n^{(1)}, \n^{(2)} - \mathbf{e}_i) D(n^{(1)}, n^{(2)})}, \\
(\n^{(1)}, \n^{(2)}) \mapsto (\n^{(1)} - \e_i, \n^{(2)} + \e_i) &\text{ w.~prob.~} \frac{c n_i^{(1)} p_2(\e_i | \n^{(1)} - \e_i, \n^{(2)})}{p_1(\e_i | \n^{(1)} - \e_i, \n^{(2)}) D(n^{(1)}, n^{(2)})}, \\
(\n^{(1)}, \n^{(2)}) \mapsto (\n^{(1)} + \e_i, \n^{(2)} - \e_i) &\text{ w.~prob.~} \frac{K c n_i^{(2)} p_1(\e_i | \n^{(1)}, \n^{(2)} - \e_i)}{p_2(\e_i | \n^{(1)}, \n^{(2)} - \e_i) D(n^{(1)}, n^{(2)})},
\end{align*}
for $i, j \in \{1, 2\}$.

It remains to specify an approximation for the conditional sampling distributions $p_i( \cdot | \cdot )$.
This was done for $\tK$ and $\tW$ in \cite{Stephens00}, and for $\tTI$ in \cite{DIG4}.
A natural approach would be to modify the generator-based method of \cite{DIG4} for $\tS$, but the resulting conditional sampling distribution vanishes for types which are present in the seed bank, but not in the active population, because mergers are blocked in the seed bank.
The trunk ancestry method of \cite{Paul10} fails for the same reason.
Instead, we suggest pooling the two populations and averaging the rates of mergers and mutations.
More precisely, let $\hat{ p }_{SD}( \e_i | \n; u )$ be the approximate conditional sampling distribution of \cite{Stephens00} for $\tK$ with mutation rate $u$, and define
\begin{equation*}
\hat{ p }( \e_i | \n^{ ( 1 ) }, \n^{ ( 2 ) } ) := \hat{ p }_{SD}( \e_i | \n^{ ( 1 ) } + \n^{ ( 2 ) }; u + u' / K ),
\end{equation*}
where the mutation rate has been obtained as the ratio of the average mutation rate, $u  K / ( K + 1 ) + u' / ( K + 1 )$ and the average merger rate $K / ( K + 1 )$.

\bibliography{Bib_WFdiffusion.bib}
\bibliographystyle{alpha}
\end{document}